\documentclass[superscriptaddress]{revtex4}
\usepackage{graphicx}

\begin{document}

\title{Quantum States of Light Produced by a High-Gain Optical
Parametric Amplifier for Use in Quantum Lithography}

\author{Girish S. Agarwal}
\address{Department of Physics, Oklahoma State University, Stillwater, OK  74078}

\author{Kam Wai Chan}
\address{The Institute of Optics, University of Rochester, Rochester, New
York 14627}

\author{Robert W. Boyd}
\address{The Institute of Optics, University of Rochester, Rochester, New
York 14627}
\address{Department of Physics and Astronomy, University of Rochester, Rochester, New
York 14627}

\author{Hugo Cable}
\address{Hearne Institute for Theoretical Physics, Department of Physics and Astronomy, Louisiana
State University, Baton Rouge, Louisiana 70803-4001}

\author{Jonathan P. Dowling}
\address{Hearne Institute for Theoretical Physics, Department of Physics and Astronomy, Louisiana
State University, Baton Rouge, Louisiana 70803-4001}
\address{Institute for Quantum Studies, Department of Physics, Texas A\&M Univerity,
College Station, TX 77843.}

\begin{abstract}
We present a theoretical analysis of the properties of an unseeded optical parametic amplifier (OPA)
used  as the source of entangled photons for applications in quantum lithography.  We first study the
dependence of the excitation rate of a two-photon absorber on the intensity of the light leaving
the OPA.  We find that the rate depends linearly on intensity only for output beams so weak that they
contain fewer than one photon per mode.  We also study the use of an $N$-photon absorber for
arbitrary $N$ as the recording medium to be used with such a light source.  We find that the
contrast of the interference pattern and the sharpness of the fringe maxima tend to increase with
increasing values of $N$, but that the density of fringes and thus the limiting resolution does not
increase with $N$. We conclude that the output of an unseeded OPA exciting an
$N$-photon absorber provides an attractive system in which to perform quantum lithography.

\end{abstract}

\date{\today }

\maketitle

\section{Introduction}
Several years ago, Boto et al.~\cite{boto00} proposed the use of entangled states of light to
produce interference patterns with sub-Rayleigh periods for use in optical lithography.  The
successful implementation of this idea could lead to many useful applications, including the
fabrication of computer chips with small feature sizes, and more generally to the development of
imaging systems that are not limited by the Rayleigh criterion. Despite the success of
proof-of-principle experiments \cite{Shih-01,Bentley} that demonstrate certain features of the
quantum lithography process, to date no true demonstration of the quantum lithography protocol has
been given. One of the major difficulties in the laboratory implementation of quantum lithography is
the conflicting requirements that the source of entangled photons be sufficiently strong to produce
multiphoton excitation of the lithographic process, yet be sufficiently weak that the statistics of
the source be essentially that of individual photon pairs.  For example, if four photons from two
independent photon pairs fall simultaneously onto the recording medium, it would be possible to
absorb one photon from each of the pairs, and this process would not lead to the the correct
sub-Rayleigh fringe pattern as envisioned by Boto et al.

It has been suggested \cite{GSA-comment,EMN-PRA,EMN-JMO} that a way to overcome this difficulty is
to replace the optical parametric downconverter envisioned in the proposal of Boto et al.~with a
high gain optical parametric amplifier (OPA) operating with a quantum vacuum input.  These papers
show by explicit calculation that the output of such a device can be arbitrarily intense yet
possess strong quantum features. In particular, in the high-gain limit the fringe visibility of
the resulting excitation pattern will become reduced, but never falls below a visibility of 20\%,
which is believed to be large enough for many practical applications.

In the present paper, we continue the analysis of the use of a high-gain OPA for use in quantum
lithography by addressing two specific questions related to the optimization of the performance of
the quantum lithography protocol.  First, we consider the case in which the recording medium is a
two-photon absorber, and we examine how the rate of excitation depends on the gain of the OPA and
hence on the intensity of the incident light.  This question is of interest because it is known
\cite{Banacloche,Javanainen,Perina,Georgiades,Silberberg} that two-photon absorption rates can scale
linearly with intensity when the optical field displays certain non-classical  features.  For
example, for a field comprised of biphotons, the two photons will tend to arrive simultaneously at a
given point on the recording material, thus leading to a two-photon absorption rate that is linear in
the intensity.  Indeed, it was suggested by Boto et al.~that this linear dependence could lead to
increased excitation efficiency which would  simplify the task of implementing quantum lithography in
the laboratory.  In the present paper we derive explicit relations that show when the excitation rate
will be linear and when it will be quadratic in the light intensity.

In this paper we also examine the situation in which the light source is
again the output of an unseeded OPA but in which the lithographic material operates by $N$-photon
absorption for arbitrary order $N$. We find that under certain circumstances the fringe
visibility is enhanced through use of a large value of $N$.  However, we find that the fringe
spacing and thus the limiting resolution is unaffected by the order $N$ of the multiphoton
absorption.

To put the ensuing theoretical development in a practical context, we next review briefly the
original quantum lithography proposal of Boto et al. As shown in Fig.~1, the
method entails generating entangled photon pairs through parametric down conversion in a nonlinear
crystal, combining these waves at a symmetric beamsplitter, and interfering these beams at a
two-photon absorbing medium.  Quantum interference effects require the two photons of
the photon pair to emerge either both in the upper arm ($\hat b_2$) or both in the lower arm
($\hat a_2$), but never one photon in each arm \cite{HOM}.  The two-photon excitation rate scales
as the square of modulus of the sum of the the probability amplitudes for
two-photon absorption for the light passing through each of the arms, leading to a fringe pattern
based on quantum interference of the form $1+ \cos 2\chi$,  where $\chi$
is the classical (one-photon) phase difference between the paths.  In contrast, the classical
interference pattern has the form $1+ \cos \chi$.

It should be noted that detection by means of two-photon absorption is now routinely used for many
optical measurements \cite{alex,west,roth}.  It should also be noted that recent work has
demonstrated the feasibility of recording interference fringes based on multiphoton excitation in
lithographic materials \cite{YAV,YAV1,Hye}.

\section{Calculation of the Two-Photon, Quantum-Lithographic Excitation Rate}

We now develop a more detailed theoretical description of the quantum lithography process. We consider
two light fields
$\hat{a}_{1}$  and
$\hat{b}_{1}$ that are generated by the process of optical parametric amplification (OPA).  Under
general
circumstances, the field operators describing these light fields can be related to those of the input
light fields $\hat{a}_{0}$  and
$\hat{b}_{0}$ by means of the relations
\begin{eqnarray}
\hat{a}_{1} =U \hat{a}_{0} + V \hat{b}_{0}^{\dag}, \\
\hat{b}_{1}=U \hat{b}_{0} + V \hat{a}_{0}^{\dag}.
\end{eqnarray}
Within the context of the present paper, we assume that the input fields are in the vacuum
state.  The coefficients $U$ and $V$ describe the strength of the nonlinear
coupling.   For parametric amplification these coefficients have the form
\begin{eqnarray} U &=& \mbox{cosh}\, G , \label{eq:uvpdc1} \\
V &=& -i\exp(i \varphi)\; \mbox{sinh}\,G, \label{eq:uvpdc2}
\end{eqnarray}
where $G$ represents the single-pass gain of the process and $\varphi$ is a
phase shift describing
the interaction. The gain factor
$G$ may be written as $G = g |E_{p}| L$ where $L$ is the length of the interaction region, $|E_{p}|$
is the pump laser amplitude, and $g$ is a gain coefficient proportional to the
second-order susceptibility
$\chi^{(2)}$.

We assume that these two generated fields are combined at a 50/50 beamsplitter. We describe the
beamsplitter by means of the standard transfer relations
\begin{equation}
\label{eq:bsgen1}
\hat{a}_{2} = \frac{1}{\sqrt{2}} \left[-\hat{a}_{1}+i\hat{b}_{1}\right]
\end{equation}
\begin{equation}
\label{eq:bsgen2}
\hat{b}_{2} = \frac{1}{\sqrt{2}}\left[i \hat{a}_{1}-\hat{b}_{1}\right].
\end{equation}
The fields leaving the beam splitter can then be
expressed as
\begin{equation}
\hat{a}_{2} = \frac{-1}{\sqrt{2}}\left[(U \hat{a}_{0} + V
\hat{b}_{0}^{\dag}) -i(U \hat{b}_{0} + V \hat{a}_{0}^{\dag})\right]
\label{eq:a2}
\end{equation}
and
\begin{equation}
\hat{b}_{2} = \frac{-1}{\sqrt{2}}\left[-i(U \hat{a}_{0} + V
\hat{b}_{0}^{\dag})
+(U \hat{b}_{0} + V \hat{a}_{0}^{\dag})\right] .
\label{eq:b2}
\end{equation}
The intensity of the light in each of these channels is then found to be given by
\begin{equation}
I = \langle\hat{a}_{2}^{\dagger}\hat{a}_{2}\rangle =
\langle \hat{b}_{2}^{\dagger}\hat{b}_{2}\rangle = |V|^{2} = \mbox{sinh}^2\,G,
\end{equation}
where we have made use of the assumption that the input fields to the OPA are in their vacuum
states.

Through use of Eqs.~(\ref{eq:a2})  and (\ref{eq:b2}), we find that the
field at the recording plane can be
written as
\begin{equation}
\hat{a}_{3}  =  \frac{1}{\sqrt{2}}\left[(-e^{i\chi}+i)(U\hat{a}_{0} + V
\hat{b}_{0}^{\dag})+ (i e^{i\chi}-1)(U\hat{b}_{0} + V
\hat{a}_{0}^{\dag})\right].
\label{field-at-medium}
\end{equation}
where $\chi$ is the classical phase difference between the two pathways. This phase difference
varies as a function of position over the detection plane.  If we restrict our attention to the
two plane-wave modes shown in Fig.~1, this phase difference can be expressed as $\chi = 2 kx
\sin \theta $  where $k= 2 \pi/ \lambda$, $\lambda$ is the fundamental wavelength associated with
each mode,
$\theta$ is the common angle of incidence of the two beams onto the recording plane,  and  $x$ is the
transverse coordinate in this plane.  We next calculate the two-photon  absorption rate at the image
plane. We express this rate as
\begin{equation}
R^{(2)} = \sigma^{(2)}
\langle\hat{a}_{3}^{\dagger}\hat{a}_{3}^{\dagger}\hat{a}_{3}\hat{a}_{3}\rangle ,
\end{equation}
where $\sigma^{(2)}$ is a generalized two-photon excitation cross section.
Again assuming a vacuum-state input to the interaction region, we find that the field-dependent
part of this rate is given by
\begin{equation}
\langle\hat{a}_{3}^{\dagger}\hat{a}_{3}^{\dagger}\hat{a}_{3}\hat{a}_{3}\rangle  =
 4|V|^{2} \left [|U|^{2}\cos^{2}\chi + 2 |V|^{2} \right ].
\label{TP-rate}
\end{equation}

We are now in a position to calculate the scaling law of the excitation rate of the lithographic
pattern.  In fact, we see from Eq.~(\ref{TP-rate}) that the scaling law is different for the
maxima and for the minima of this pattern.  At the minima of the pattern, where $\cos^{2}\chi
=0$, the excitation rate is given by
\begin{equation}
R^{(2)}_{\rm min} = 8 \sigma^{(2)} |V|^4 = 8 \sigma^{(2)} \mbox{sinh}^4\,G = 8 \sigma^{(2)} I^2.
\end{equation}
We thus find that at the minima of the fringe pattern the two-photon excitation rate always scales
quadratically with intensity.  However, at the maxima of the fringe pattern, where $\cos^{2}\chi
=1$, we find that the excitation rate is given by
\begin{equation}
R^{(2)}_{\rm max} = 4 \sigma^{(2)} |V|^2 \left[ |U|^2 + 2|V|^2 \right ] = 4 \sigma^{(2)}
\mbox{sinh}^2\,G  \left [\mbox{cosh}^2\,G  + 2 \mbox{sinh}^2\,G \right ]  = 4 \sigma^{(2)} (I + 3
I^2) .
\end{equation}
Thus, we find that at the fringe maxima the excitation rate has both a linear and a
quadratic contribution.  This
behavior is illustrated in Fig. 2.   The cross-over point between the linear and quadratic behavior
occurs at an intensity of approximately $I = 1/3$ photons per mode or a gain coefficient of $G =
0.55$.  We thus conclude that for essentially all cases of practical interest the  excitation rate
scales quadratically with intensity.  It should be noted, however, that this conclusion follows only
for the case of the output of an OPA.  For other states of light, such as the pure biphoton state,
the linear scaling relation would hold for arbitrarily large intensities. It is also worth noting
that for all values of the intensity $I$, the spatially varying part of the excitiation pattern
oscillates at twice the spatial frequency of the classical interference pattern.  The only
consequence of the use of a large intensity $I$ is to induce a uniform background upon which the
fringe pattern sits. Thus increased spatial resolution is obtained even for large values of $G$ and
$I$.

\section{Quantum Lithography with an $N$-photon recording material}

We now consider the situation in which the recording material operates by $N$-photon absorption for
an arbitrary value of the order $N$.  As before, the field at the recording medium is given by
Eq.~(\ref{field-at-medium}).  The $N$-photon absorption rate can be expressed as
\begin{equation}
R^{(N)} = \sigma^{(N)}
\langle\hat{a}_{3}^{\dagger N}\hat{a}_{3}^N\rangle ,
\end{equation}
where $\sigma^{(N)}$ is a generalized $N$-photon excitation cross section.  The field-dependent part
of this quantity can be straightforwardly evaluated and is given by
\begin{equation}
    \langle  \hat{a}_3^{\dagger N} \hat{a}_3^N \rangle
    = \sum_{n=0}^{\lfloor N/2 \rfloor} 2^{N-2n}
    \left|P_{N-2n}^{N}\right|^2 |V|^{2(N-n)} |U|^{2n} \cos^{2n}\chi .
\end{equation}
where the quantity $P_{m}^{n}$ is given by the recurrence relation
\begin{eqnarray}
&&  P_{N}^{N}
    = \sqrt{N!} , \qquad\text{and}
    \\
&&  P_{N-2n}^{N}
    = 2\sqrt{(N-2n+1)} P_{N-2n+1}^{N-1} + \sqrt{(N-2n)} P_{N-2n-1}^{N-1} ,
    \quad\text{where $n=0,1,\cdots , \left\lfloor N/2 \right\rfloor$.}
\end{eqnarray}
For example, we find using these formulas that the four lowest-order multiphoton absorption rates are
given by
\begin{eqnarray}
R^{(2)} = \sigma^{(2)} \langle  \hat{a}_3^{\dagger 2} \hat{a}_3^2 \rangle
 &=&   \sigma^{(2)}
    4 |V|^{2} \Big( 2 |V|^{2}
    + |U|^{2} \cos^{2}\chi \Big) ,
\\
R^{(3)} =  \sigma^{(3)} \langle  \hat{a}_3^{\dagger 3} \hat{a}_3^3 \rangle   &=&
\sigma^{(3)}
    24 |V|^{4} \Big( 2 |V|^{2}
    + 3 |U|^{2} \cos^{2}\chi \Big) ,
\\
R^{(4)} = \sigma^{(4)} \langle  \hat{a}_3^{\dagger 4} \hat{a}_3^4 \rangle
 &=& \sigma^{(4)}
    48 |V|^{4} \Big( 8 |V|^{4}
    + 24 |V|^{2} |U|^{2} \cos^{2}\chi
    + 3 |U|^{4} \cos^{4}\chi \Big) ,
\\
R^{(5)} = \sigma^{(5)} \langle  \hat{a}_3^{\dagger 5} \hat{a}_3^5 \rangle
 &=& \sigma^{(5)}
    480 |V|^{6} \Big( 8 |V|^{4}
    + 40 |V|^{2} |U|^{2} \cos^{2}\chi
    + 15 |U|^{4} \cos^{4}\chi \Big) .
\end{eqnarray}
Defining the $N$-photon fringe visibility as
\begin{equation}
    V(N) = \frac{R_{\max}^{(N)} - R_{\min}^{(N)}}{R_{\max}^{(N)} + R_{\min}^{(N)}} ,
\end{equation}
we can readily calculate the dependence of $V(N)$ on the gain $G$, as shown in
Fig.~\ref{fig:visibility}. We see that through use of large values of $N$ the fringe visibility
remains large as the gain $G$ is increased.   However, Figs.~\ref{fig:AbsRate-G=0.1},
\ref{fig:AbsRate-G=0.5} and~\ref{fig:AbsRate-G=1} show that while high-order multiphoton absorption
can produce narrower fringes, the fringe spacing remains the same.  We also see that narrowed
fringes occur only when the gain $G$ is less than unity.

\section{Summary and Conclusions}

In summary, we have developed a theoretical model that describes how the output of an unseeded
parametric amplifier can be used as the source of entangled photons to be used to perform quantum
lithography.  We have used this model first to determine the excitation rate for a quantum
lithographic material that operates by means of two-photon absorption.  We find that in general the
transition rate has two contributions, one of which is linear and the other of which is quadratic in
the light intensity.  We also find that the linear term dominates only for very weak beams of light
that contain on average far fewer than one photon per mode.  Since beams this weak are unlikely to
prove useful in the context of quantum lithography, we conclude that under all practical situations
the quadratic term is expected to dominate.  At one time it had been hypothesized that it would be
desirable to perform quantum lithography under conditions of linear response \cite{boto00}.   This
hypothesis was based on the argument that a linear response would increase with excitation strength
more rapidly than a quadratic response under conditions of low excitation.  The present analysis
shows that one can easily work under conditions such that the quadratic, more-rapidly-growing term
dominates. We have also examined the use of such a light source in the context of a lithograpic
recording medium that operates by means of $N$-photon absorption for arbitrary $N$.  We find that the
use of a large value of $N$ allows the fringe visibility to remain large even for moderately large
values of the OPA gain $G$.  However, the use of large $N$ does not lead to an increase of the fringe
density and hence of the spatial resolution of the lithographic process.  The use of a large
nonlinear order $N$ does lead to narrower fringes, but only for values of the gain $G$ that are less
than unity.   On the arguments presented in this paper, we conclude that an unseeded OPA exciting an
$N$-photon absorber provides an attractive system with which to perform quantum lithography.

\bigskip

The portion of the work conducted at the University of Rochester and at Louisiana State University
was supported by the US Army Research Office under a MURI award. KWC gratefully acknowledges
support by the Croucher Foundation.  GSA thanks NSF for supporting this work through grant
CCF-0524673.  JPD and HC would also like to acknowledge support from the National Security Agency,
the Distruptive Technologies Office, and the Hearne Institute for Theoretical Physics.

\newpage

\begin{figure}[!ht]
\includegraphics[scale=0.5]{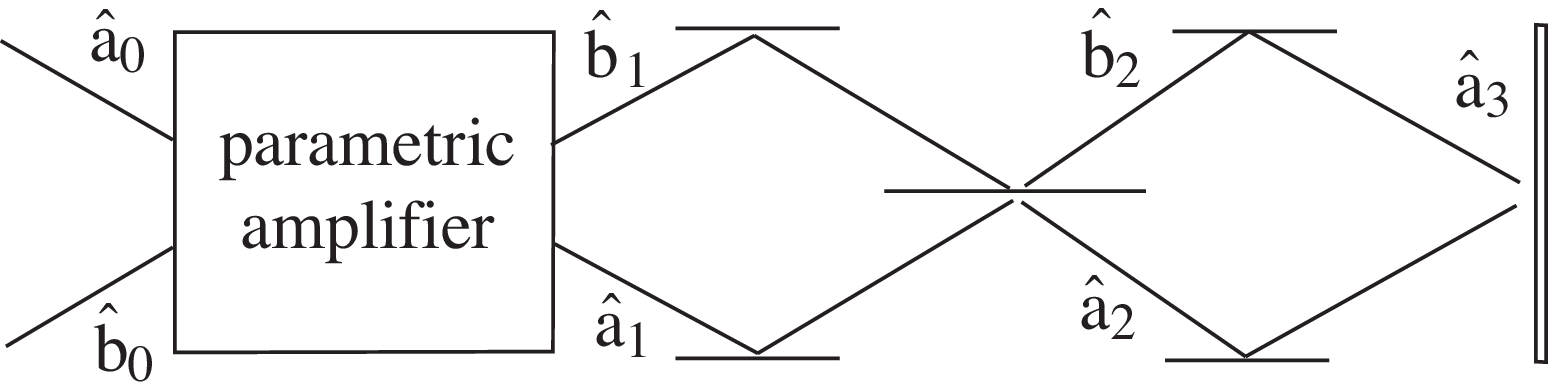}
\caption{Schematic representation of the quantum lithography architecture. }
\end{figure}

\begin{figure}[!ht]
\includegraphics[scale=0.5]{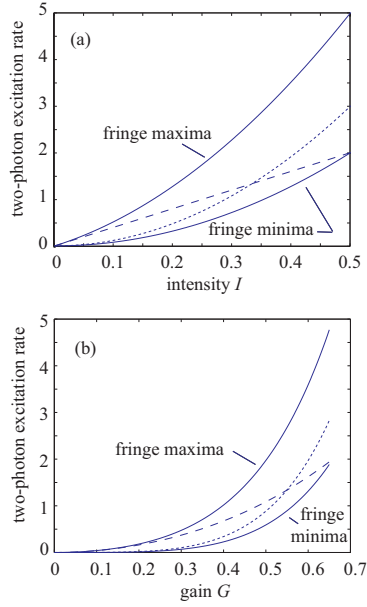}
\caption{(a) Two-photon excitation rate for the fringe maxima and minima plotted as functions of
the intensity in each of modes $\hat a_2$ and $\hat b_2$.  The dashed and dotted curves show
respectively the linear and quadraic contributions to the excitation rate for the fringe maxima.  (b)
Same as part (a), but plotted as functions of the OPA gain $G$. }
\end{figure}

\begin{figure}[!h]
\includegraphics[scale=0.5]{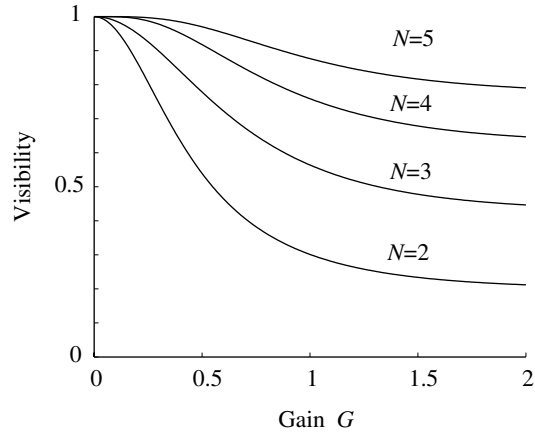}
\caption{Fringe visibility $V(N)$ plotted as a function of the gain $G$ for various values of the
order $N$ of the multiphoton absorption process.}
\label{fig:visibility}
\end{figure}

\begin{figure}[!h]
\centering\includegraphics[height=6.5cm]{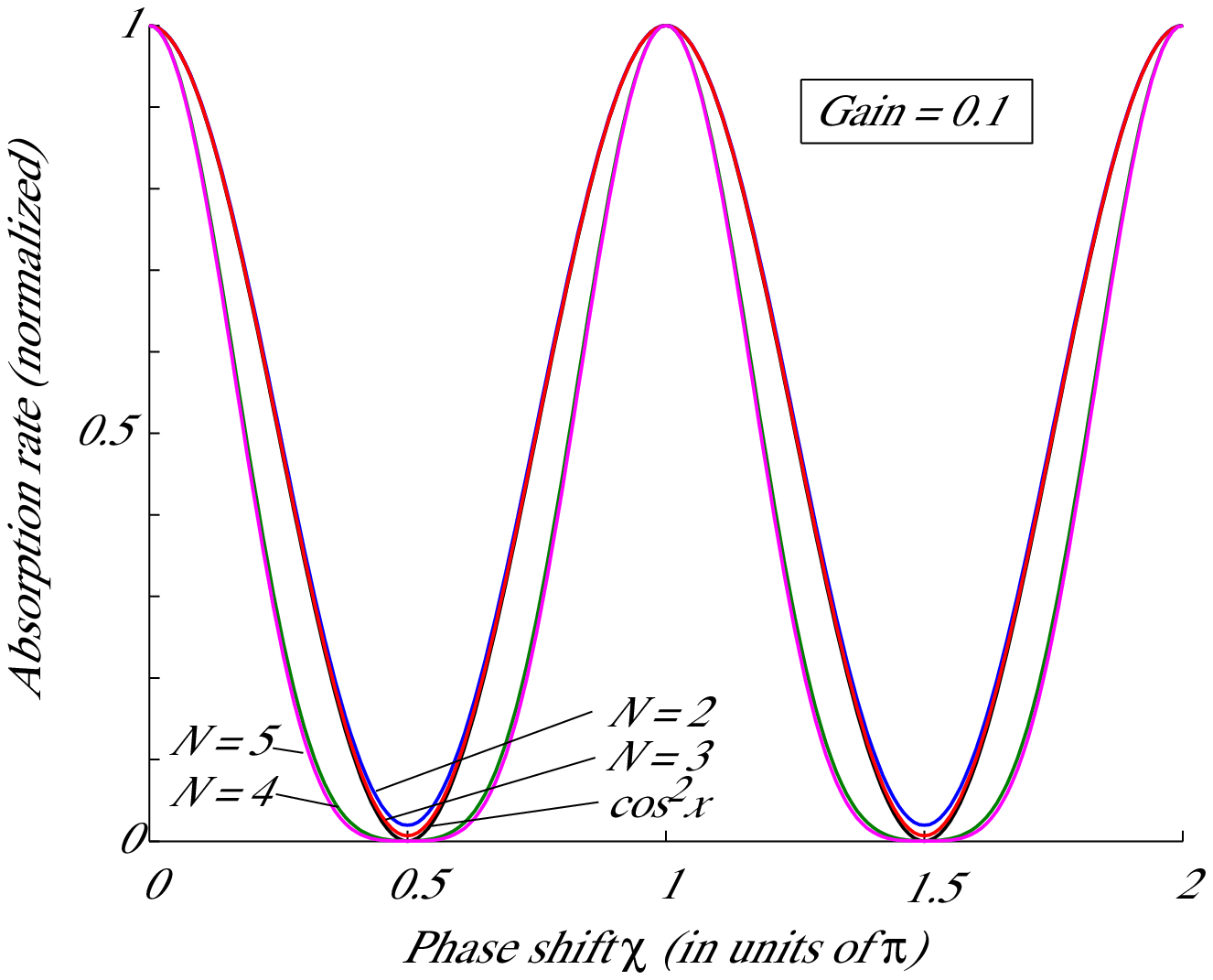}
\caption{Absorption rate $R^{(N)}$ plotted as a function of the classical phase shift $\chi$
for a  gain of $G=0.1$.}
\label{fig:AbsRate-G=0.1}
\end{figure}

\begin{figure}[!h]
\centering\includegraphics[height=6.5cm]{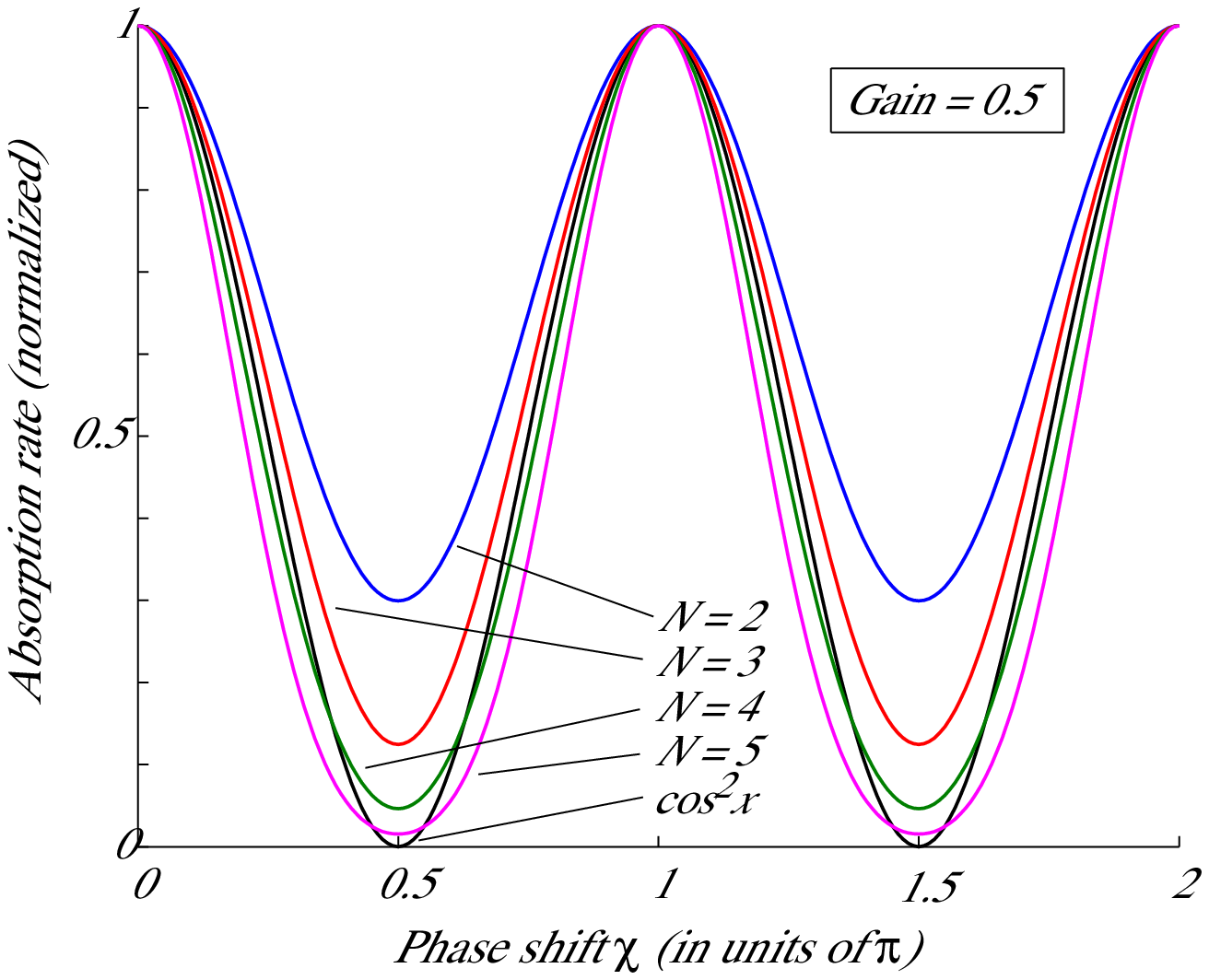}
\caption{Absorption rate $R^{(N)}$ plotted as a function of the classical phase shift $\chi$
for a  gain of $G=0.5$.}
\label{fig:AbsRate-G=0.5}
\end{figure}

\begin{figure}[!h]
 \centering\includegraphics[height=6.5cm]{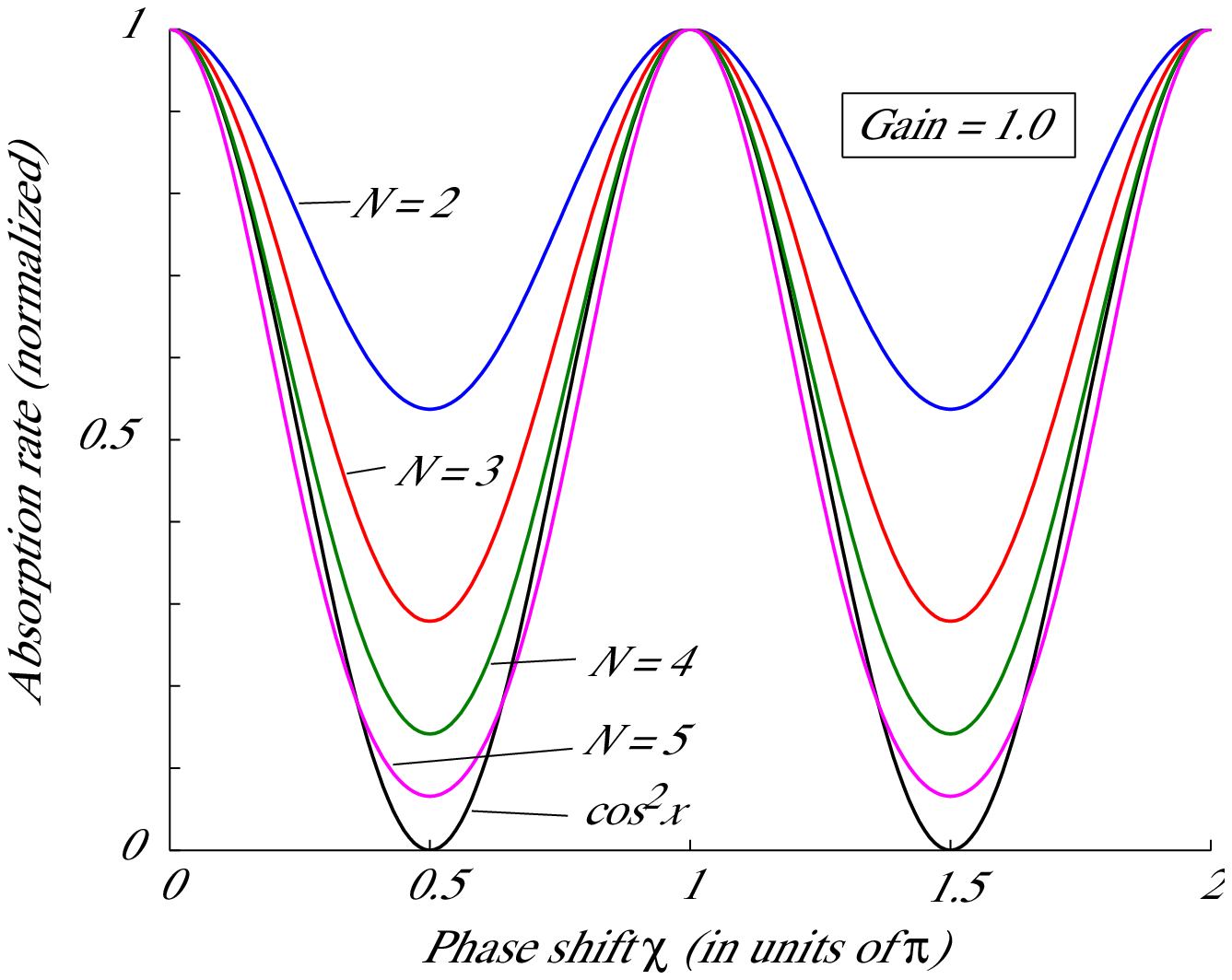}
\caption{Absorption rate $R^{(N)}$ plotted as a function of the classical phase shift $\chi$
for a  gain of $G=1.0$.}
\label{fig:AbsRate-G=1}
\end{figure}

\end{document}